# Design and Implementation of Domain based Semantic Hidden Web Crawler

**Manvi**
Department of Computer Engineering
YMCA University of Science & Technology
Faridabad, India

**Ashutosh Dixit**
Department of Computer Engineering
YMCA University of Science & Technology
Faridabad ,India

**Komal Kumar Bhatia**
Department of Computer Engineering
YMCA University of Science & Technology
Faridabad, India

**Jyoti Yadav**
Department of Computer Engineering
YMCA University of Science & Technology
Faridabad, India

**Abstract:-**Web is a wide term which mainly consists of surface web and hidden web. One can easily access the surface web using traditional web crawlers, but they are not able to crawl the hidden portion of the web. These traditional crawlers retrieve contents from web pages, which are linked by hyperlinks ignoring the information hidden behind form pages, which cannot be extracted using simple hyperlink structure. Thus, they ignore large amount of data hidden behind search forms. This paper emphasizes on the extraction of hidden data behind html search forms. The proposed technique makes use of semantic mapping to fill the html search form using domain specific database. Using semantics to fill various fields of a form leads to more accurate and qualitative data extraction.

## 1. INTRODUCTION

World Wide Web is becoming an important source of information these days. It comprises of Surface web and hidden web [1]. Surface web refers to the part of web which we can access via hyperlinks or predefined URL's. Hidden Web (also called the Deep net, Invisible Web, or Deep Web) refers to the information that is "hidden" behind the query interfaces.

Hidden Web is becoming very important as it stores very large amount of high quality information. Recent studies have estimated the size of this hidden web at around 500 times the size of Publicly Indexed Web [2][4]. As the volume of hidden information is growing at a fast pace, there has been increased interest to develop techniques that can allow users and applications to access this information.

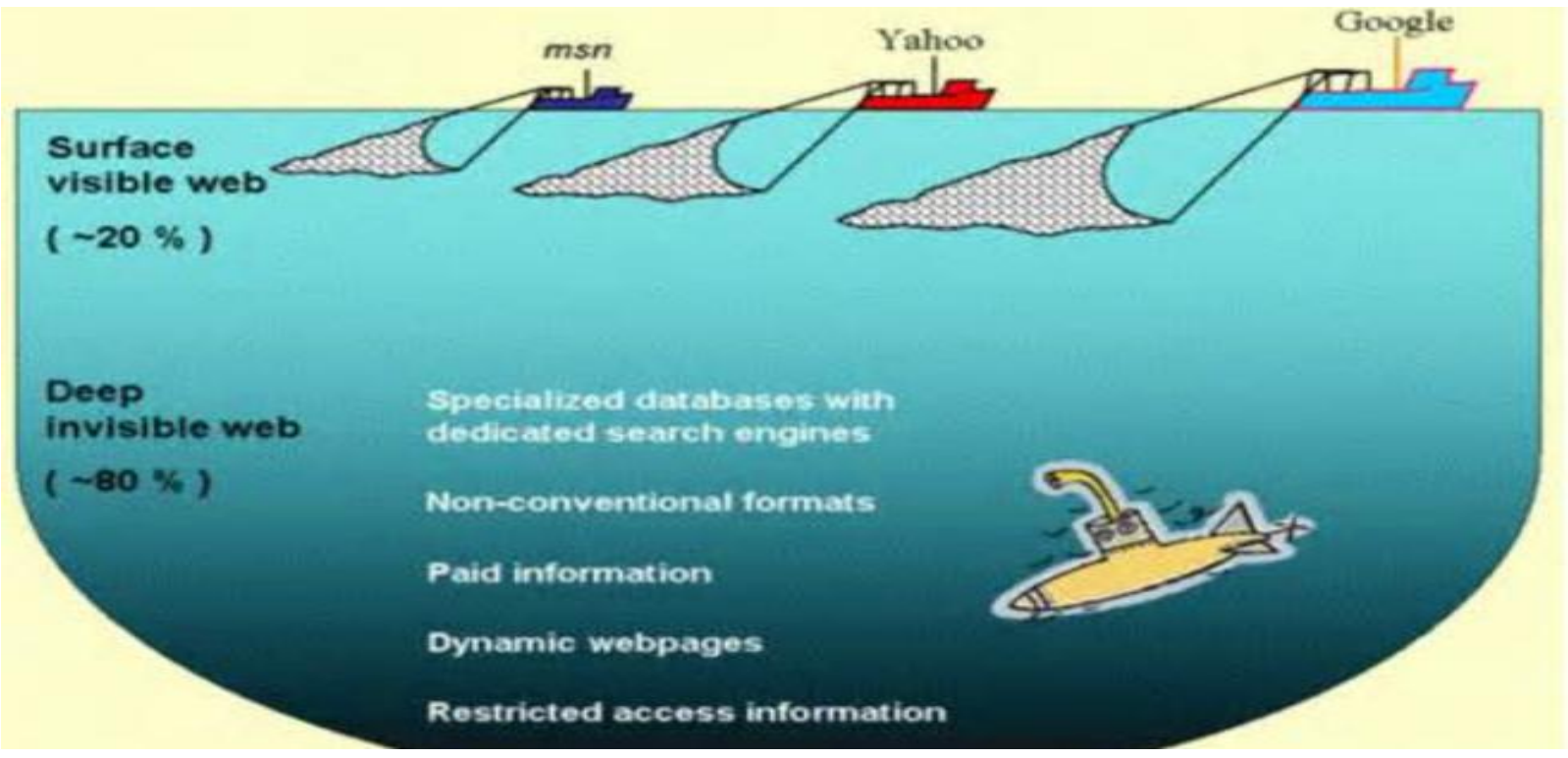

Fig. 1 Surface Web and Hidden Web.

Simple Web crawlers discover the various web pages on basis of hyperlinks available on web pages [3].These web crawlers are not capable of discovering and indexing the hidden web pages as there are no static links to them. The large amount of high quality information is obscured under dynamically generated web pages and the only entry point to access this information is through various query interfaces.

So to access hidden web data one needs to fill various search forms available on different websites. There are two ways to fill these field values First is that a user manually sits on each website fill the required information and get the result, the other way is to design a specialized crawler which automatically fill these values and to automatically fill forms crawler must have accurate data as datasets or databases.

These datasets/databases may be created manually or automatically using various techniques. Ontology may be one of them as this will provide meaningful, accurate and relevant information of data values. Ontology itself is a vast field and requires a lot of understanding. Making Ontology of all domains is not possible in one go. So in this paper extraction of hidden web data for specific book domain is done. The crawler here is being guided by a set of data values based on the domain of interest and crawls the web focusing on pages relevant to a given domain.

## 2. LITERATURE SURVEY

Many researchers are trying to develop novel ideas to access hidden web in order to improve crawling techniques for Hidden web. A brief overview at few of them is given in the following subsections:

**A. HIWE by Sriram Raghavan [7]:**This work proposed a task-specific hidden-Web crawler, the focus of this work was to develop Hidden-Web query interfaces .A prototype hidden Web crawler called HiWE (Hidden Web Exposer) was developed. The first limitation is HiWE‟s inability to recognize and respond to simple dependencies between form elements (e.g., given two form element corresponding to states and cities, the values assigned to the „city‟ element must be cities that are located in the state assigned to the „state‟ element).The other limitation is HiWE‟s lack of support for partially filling out forms; i.e., providing values only for some of the elements in a form.

**B. Framework for Downloading Hidden Web Content[8]** : Ntoulas et al. differ from the previous studies, that, it provided a theoretical framework for analyzing the process of generating queries , In his work it was concluded that the only "entry" to Hidden Web pages is through querying a search form, there are two core challenges to implementing an effective Hidden Web crawler: (a) The crawler has to be able to understand and model a query interface, and (b) The crawler has to come up with meaningful queries to issue to the query interface. Raghavan and Garcia-Molina addressed the first challenge, where a method for learning search interfaces was presented. Here, they gave solution to the second challenge, i.e. how a crawler can automatically generate queries so that it can discover hidden information. This body of work was often referred to as database selection problem over the Hidden Web. The disadvantage was that the work only supported single attribute queries.

**C. Madhavan et.al. [9]** : They proposed Depth-Oriented crawler for content extraction. It extract hidden data from websites having multi-attribute search interface and having structured database at backend. It evaluate the query templates by defining the informativeness test. This technique efficiently navigates the search space of possible input combinations but there is no consideration to the efficiency of deep web crawling.

**D. Xiang Peisu et al. [10]** proposed model of forms and form filling process that concisely captures the actions that the crawler must perform to successfully extract Hidden Web contents. It described the architecture of the deep web crawler and described various strategies for building (domain, list of values) pairs. Although this work extracts some part of Hidden Web but it is neither fully automated nor scalable.

**E. Data Extraction from Hidden web Databases [6] :** This work by Anuradha et al. proposed four phased architecture for extraction of hidden web databases. Firstly, different query interfaces are analyzed to select the attribute for submission. In the second phase, queries are submitted to interfaces. Third phase extracts the data by identifying the templates and tag structures. Fourth phase integrates the data into one repository. They emphasized on different query methods to submit form such as blank query, query with all combination and query selection on mandatory fields. After submitting form data is extracted by looking at tags as data stored at backend in structured form (tables). This paper cannot extract data from websites having unstructured databases. It mainly focuses on the query submission .It lacks in semantic mapping of query while extracting data from hidden web databases.

| Techniques | Structured | Unstructured | Simple Search Interface | Classification | Query Probing | Use of Ontology |
|---|---|---|---|---|---|---|
| Raghavan and Garcia-Molina | √ | | √ | √ | | |
| Ntoulas et al. | √ | √ | √ | | √ | |
| Barbosa and Freire | √ | √ | √ | √ | | |
| Madhavan, David Ko | | √ | √ | | | |
| Anuradha et al | √ | | √ | √ | | √ |

Table.No.1. Comparison table for all the works

## 3. PROBLEM IDENTIFICATION

Although there are many proposed crawling techniques to extract data from hidden web still there are some issues which need to be addressed. By doing a critical analysis of above papers some problems are identified as-

- The available crawling techniques which give some efficient automated results supports only single attribute queries.
- The available hidden web crawler techniques cannot extract data from websites which used unstructured databases at backend.
- In available techniques there is lack of support for partially filling out forms; i.e., providing values only for some of the elements in a form.
- Another limitation is inability to recognize and respond to simple dependencies between form elements (e.g., given two form element corresponding to states and cities, the values assigned to the city‟ element must be cities that are located in the state assigned to the „state‟ element).

In order to resolve the identified problems Domain based semantic data extraction and integration is done. By using Semantic analysis multi attribute or single attribute forms can be identified and these form can be filled by identifying similar label values from the task specific database. This will make retrieval process more efficient by increasing the likelihood of being able to extract the relevant subset of data.

## 4. PROPOSED WORK

There are six main modules in the proposed Domain based Semantic Hidden Web Crawler as shown in Figure 2.

i) Website Identifier and Form Downloader
ii) Form Filler
iii) Semantic Label Identifier
iv) Result Processor
v) Redundancy Remover
vi) Updater(Feedback Module)

With these above stated modules there is a database which is also being created used and manipulated (updated) in this work. The name of the database is **Task-specific Database.** It is task oriented in the sense that it has been used to store values for a particular domain that is book domain for the proposed work.

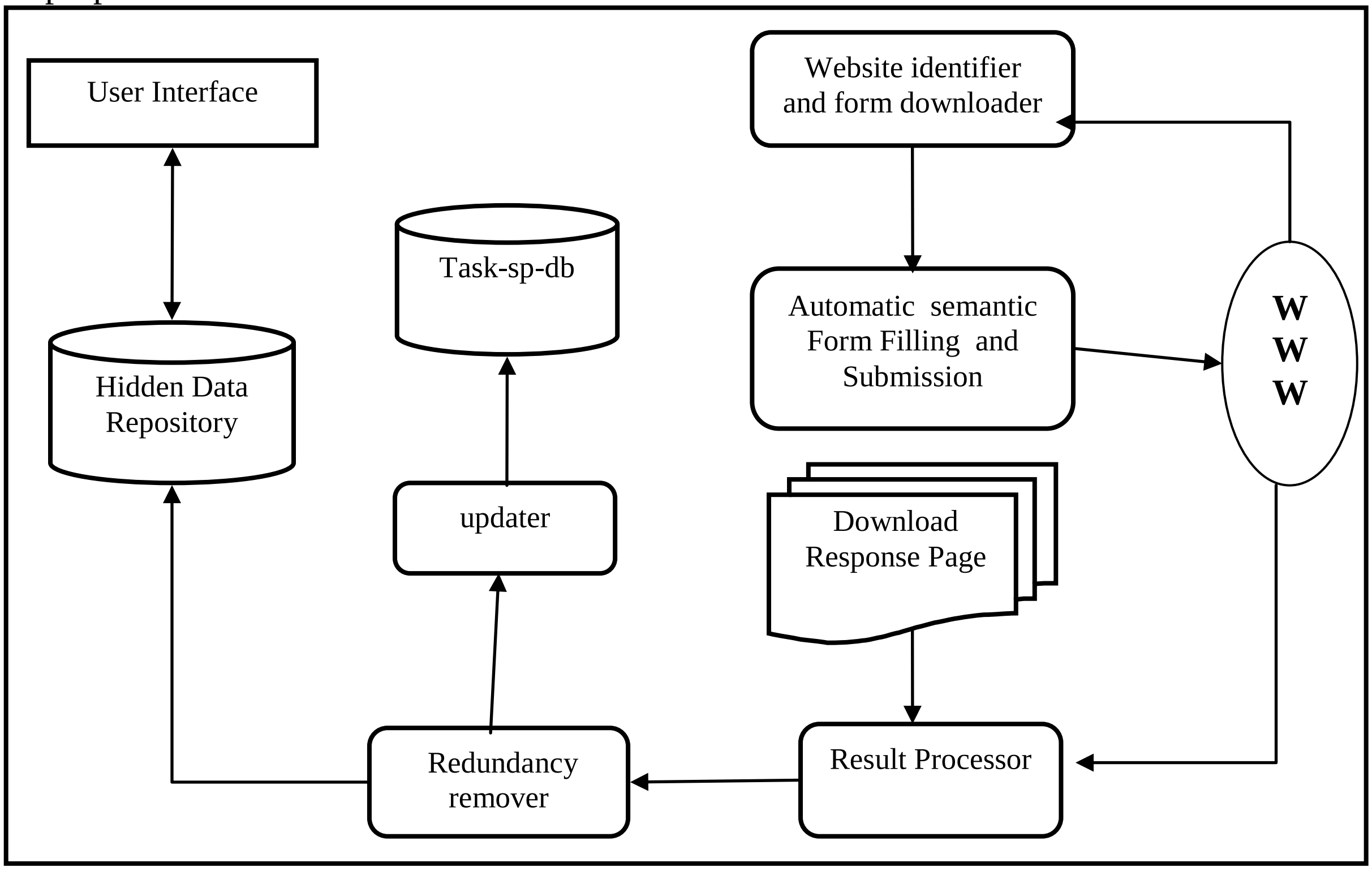

Figure 2 . Proposed architecture of Domain based Semantic Hidden Web Crawler.

Detailed explanation of all the modules is given below:

**i) Website Identifier and Form Downloader:** In this Module Hidden Web Crawler will identify the websites having any query interface (html search form) for extraction of data from Hidden Web. This search form or query interface is identified by checking HTML code of the page. If tag FORM is found in the web page then it will download from the world wide web to extract data stored at backend of website.

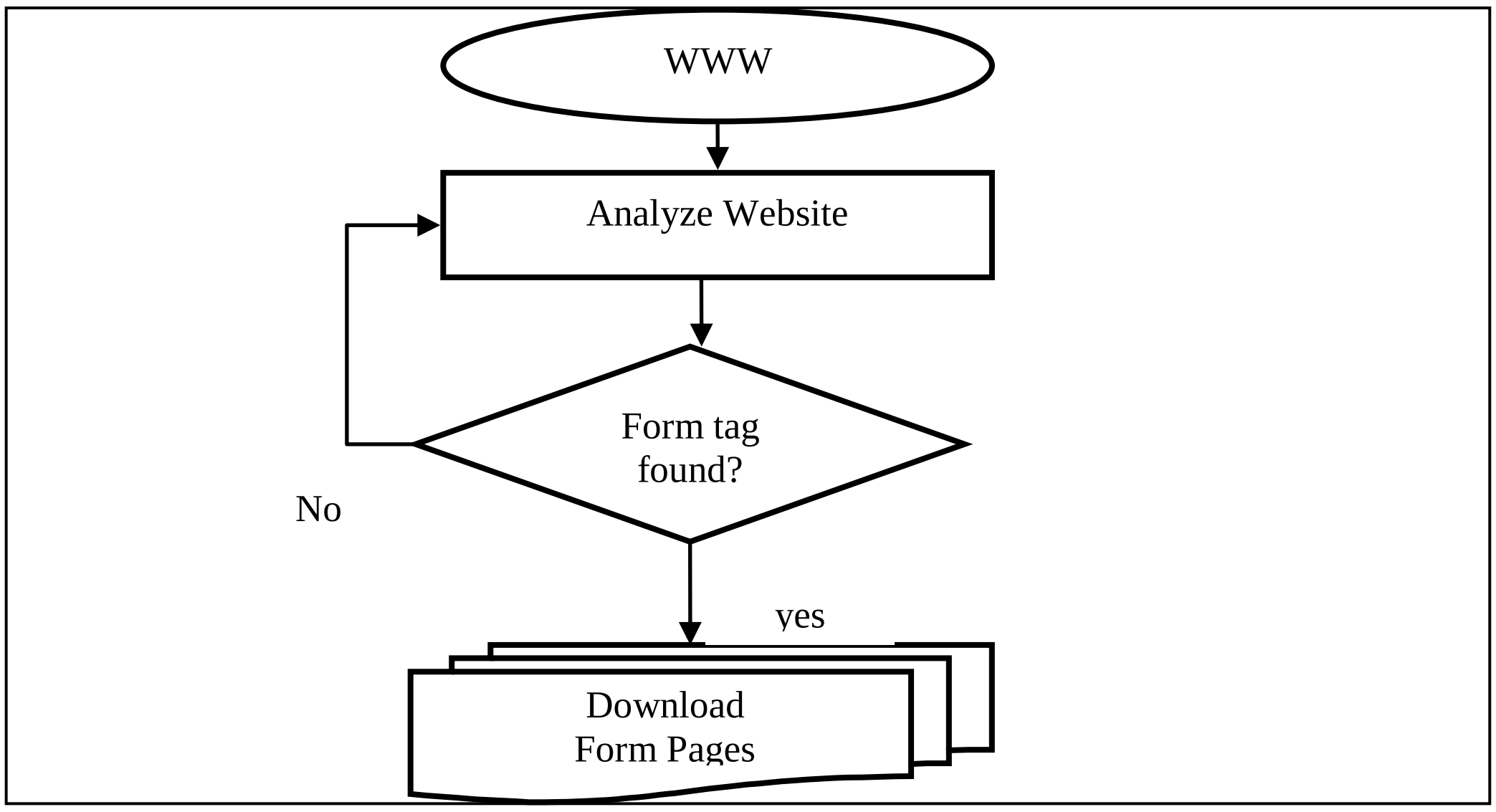

Fig 3. Website Identification and Form Download

**ii) Automatic Semantic Form Filler :** Form on the search page of any website can be of two types i.e single attribute form and multiple attribute form. Given below is the description of automatic form filler component which further consist of following sub components:

- Semantic Label Identifier(SLI)
- Label Matcher
- Automatic form filler
- Task Specific Database

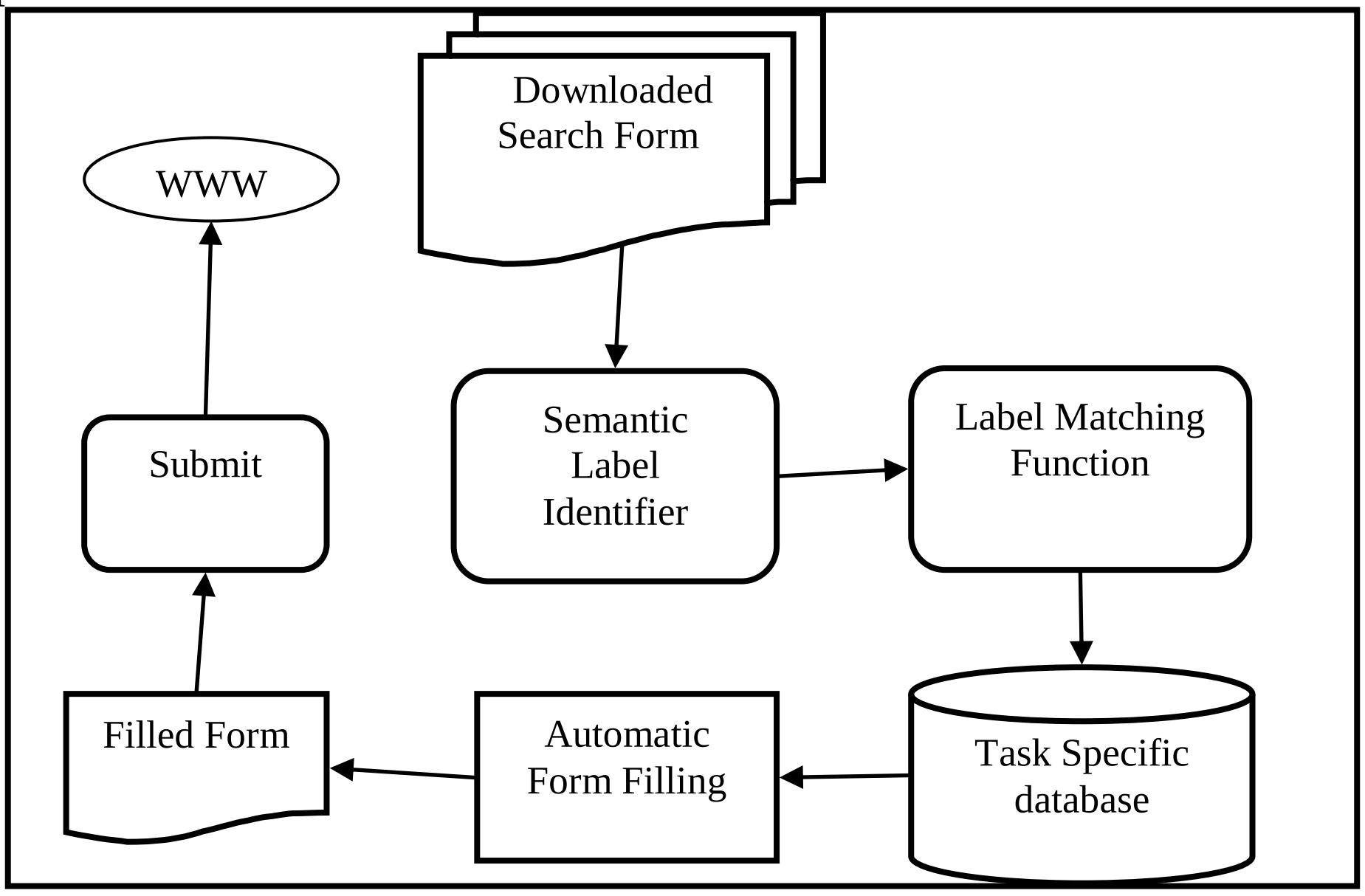

Fig 4. Form Filling

➢ **Semantic Label Identifier:** In this Sub-Module form will be analyzed and crawler will retrieve label names attached with various attributes of forms. For instance, form F = (fA1; fA2… ), fA1,fA2 are attributes, crawler will collect information about domain of attribute and its label(Li). The domain of an attribute defines the set of values which can be associated with the corresponding form attribute. Some attributes have finite domains, where the set of valid values are already embedded in the page. For example, if Aj is a selection list, then info (Aj) is the set of values that are contained in the list. Other attributes with free-form input, such as text boxes, have infinite domains (e.g., set of all text strings). The label of a form element is the descriptive information associated with that attribute. Most forms are usually associated with some descriptive text to help the user understand the semantics of the attribute.

➢ **Label Matcher:** Now this domain and label of form will be matched by label matching function with the fields of task specific database. The Label matching function identifies various synonyms or similar label names and maps them with one set of values of task –specific database. Values having equal or more than a threshold matching value can be filled from database to form. For Ex. for a query interface of book domain if form has field label names as 'Title' , 'Written by' and "Published By" then it will be matched to Label names 'Title/Subject' and 'Author' of task specific database respectively.

➢ **Automatic Form Filling And Submit :** After matching the label names and finding data for filling the form, crawler will fill the form with selected data values and submit the form to the web server.

**iii) Task-specific Database** : In Hidden web crawler, task-specific database is organized in terms of a finite set of concepts or categories. Each concept has one or more labels and an associated set of values. For example, the label 'Author Name' could be associated with the set of values 'R.P Jain', 'Morris Mano', 'Yashwant Singh'. This task specific database is created automatically using some seed urls by the crawler . It will download data from some websites by filling their forms manually and retrieving some set of results as shown in fig. 4.

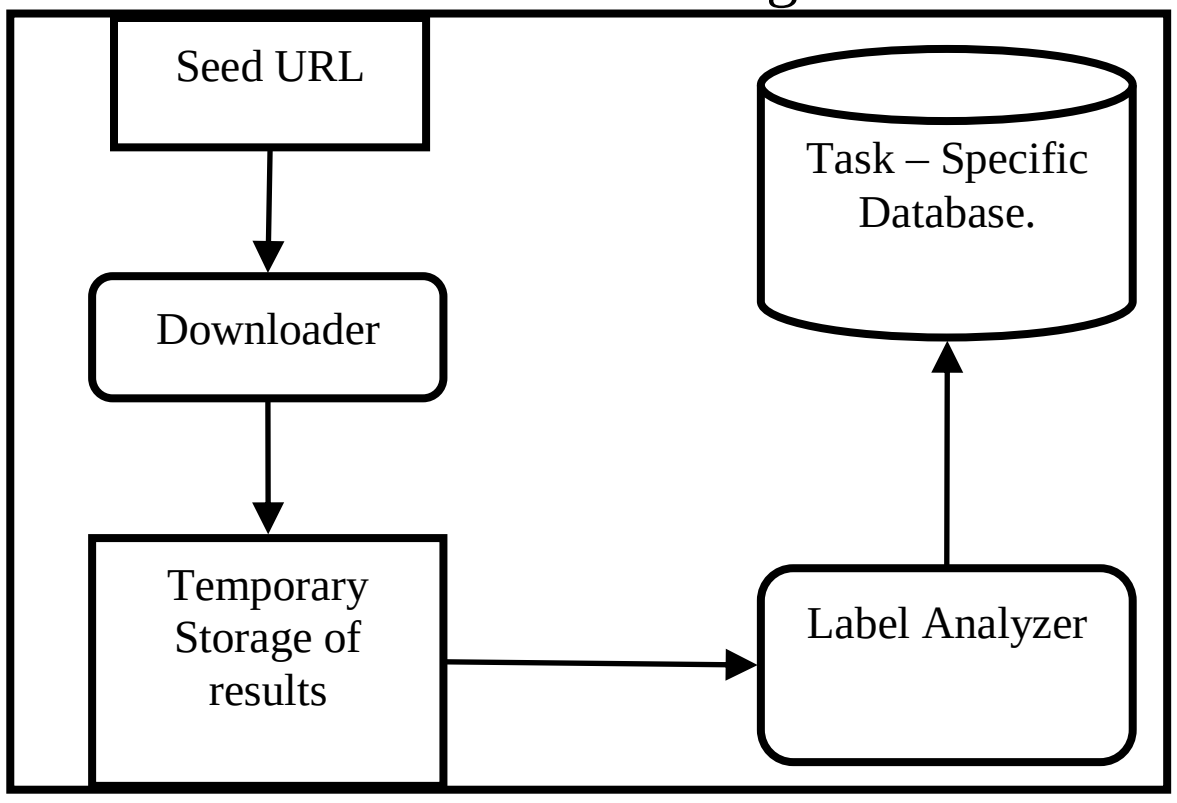

Fig. 5 Creation Of Task Specific Database.

Components used in creation of task specific database are explained as:

➢ Downloader: This component takes input in the form of the URL (initially manually) and then downloads the page corresponding to the URL in basic manner from WWW. The resultant pages are stored in a temporarily storage to provide input to next component i.e Label Analyzer.

➢ Label Analyzer: This sub module for each page, identify the labels attached and their values associated. Thus task-Specific Database is saved in the database in the form of table called as Label-Value Table for further utilization.

➢ Download Response pages: After submission of filled search form to the website it will generate response page. This response page will be downloaded from the web server.

**iv) Result Processor:** This module has two sub-modules:-

➢ **Result Retrieval:** In this Sub- Module crawler will have the response page with it which have the results of the search query embedded within this page. Now crawler will need semantic mapping based on ontological information. For e.g. response page of any book domain may have author of book by label author, written by or writer etc, all these will be matched to author label of repository. It will search and select data from the response page using information of various fields' value and meta information of the tags used in the response web page.

➢ **Integrator:** It will integrate all the results retrieved by different websites and fill ing various forms. These integrated retrieved results will be now stored in the Hidden data Repository database after removing redundancy.

**v) Redundancy Removal:** Response page may have some redundant results which user do not need. It is also required in order to separate error pages. For e.g. Any book domain response page may show book which is out of stock or not available. Response page may have repetition of same book. Such redundancies must be removed before storing the results in repository. As data is extracted from various hidden databases of websites by submitting same query to respective local interfaces. However, it is very much possible that many of these sites contain same results. Also while storing results there may come some duplication with previously stored results . Hence, data repository should be made in such a way that duplicate records are removed while merging , for better database management. **Removing Duplicate Records:** To remove duplicate records, Sql query is fired and all the duplicate results are removed from data repository.

**vi) Updater :** In this module task specific database is updated if the labels and values of retrieved results are not present in the database. It will store the corresponding labels and values in the database so that these values can be used to fill form to get more related results.

**B) Basic Flow Diagram of the entire system:** The basic working flow of the proposed system is described in the flow diagram as shown in figure 6.

In order to extract data from the websites having hidden web data following steps are required:

Step i)Website identification and Download query interface or search form.

Step ii) Filling of form using semantic mapping and Submit the form to web server.

Step iii) Download response Page generated after submission of form from web server.

Step iv) Result retrieval semantically from downloaded response pages. Check for redundant results. Store the results in local databases. Integrate the database and remove duplicate records.

The basic working flow of the proposed system is described in the flow diagram as shown in figure 6.

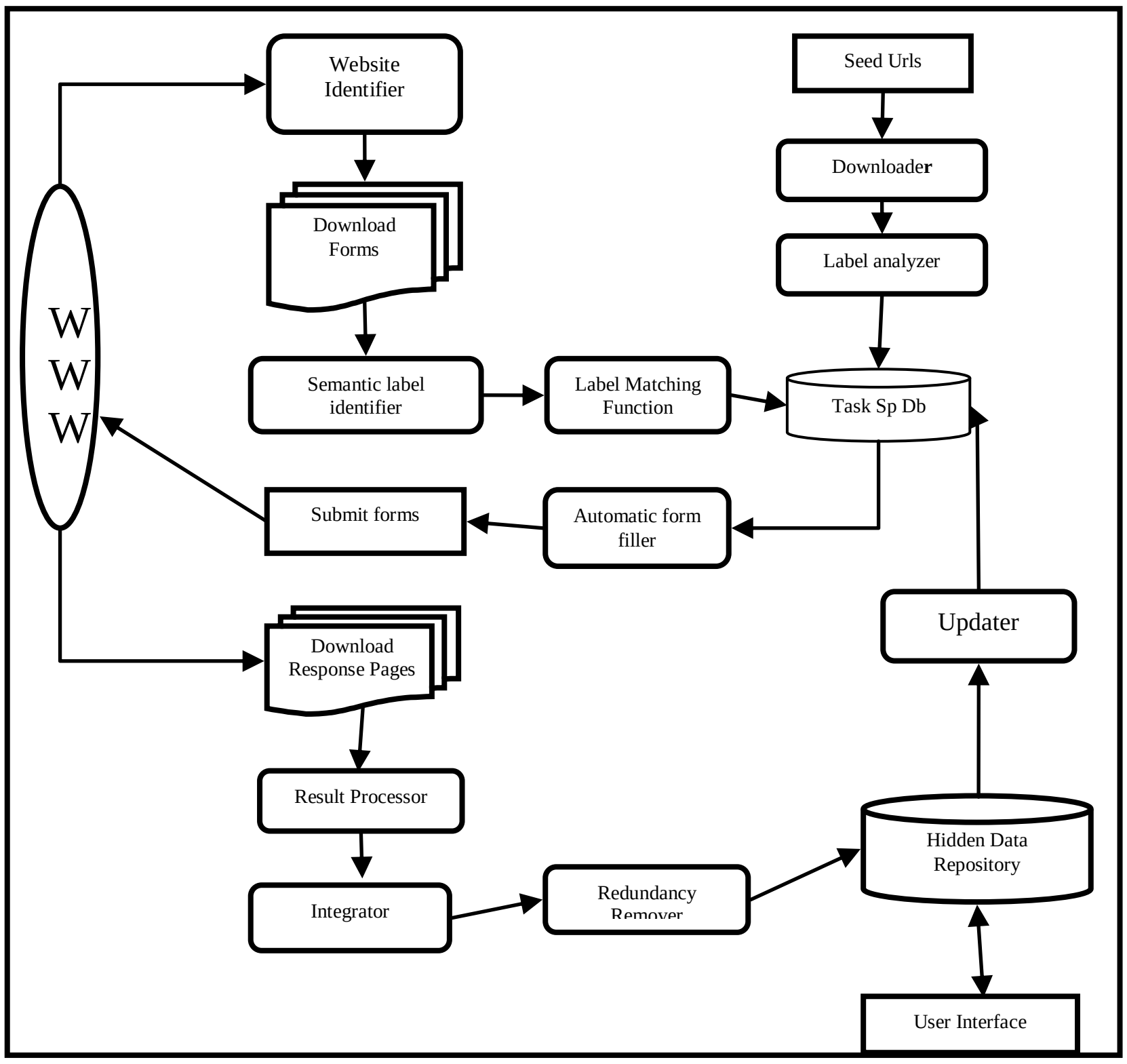

Fig.6. Flow of Proposed Work.

## 5. IMPLEMENTATION AND RESULTS

The proposed architecture is implemented for Book domain taking websites having single attribute search form and multi attribute search forms as shown in figure 7 and figure 8 respectively. Figure 7 taken from the site "www.powells.com" demonstrates those form pages which contain a single text box for user to enter queries. However figure 8 , gives the example of the website www.abebooks.com which consist of more than one text box where user fills the desired information manually to get the desired results.

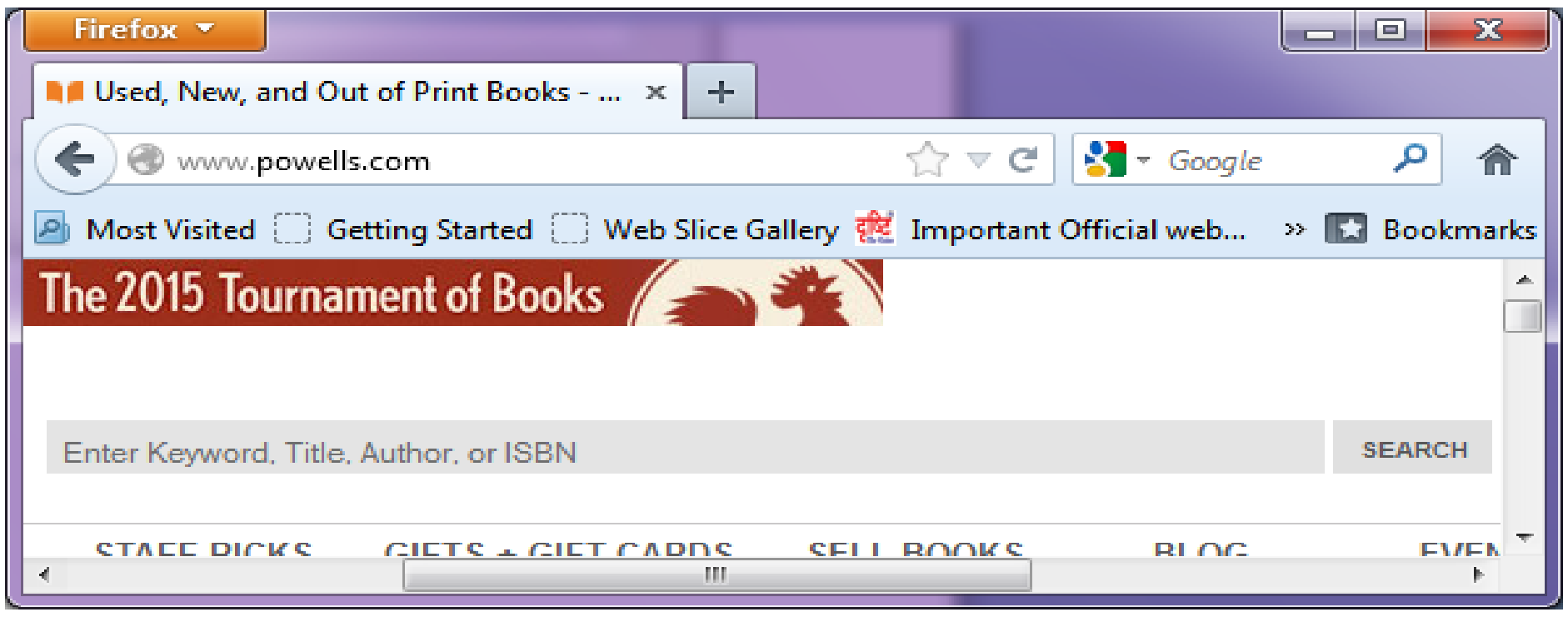

**Fig. 7** Single-attribute Search form.

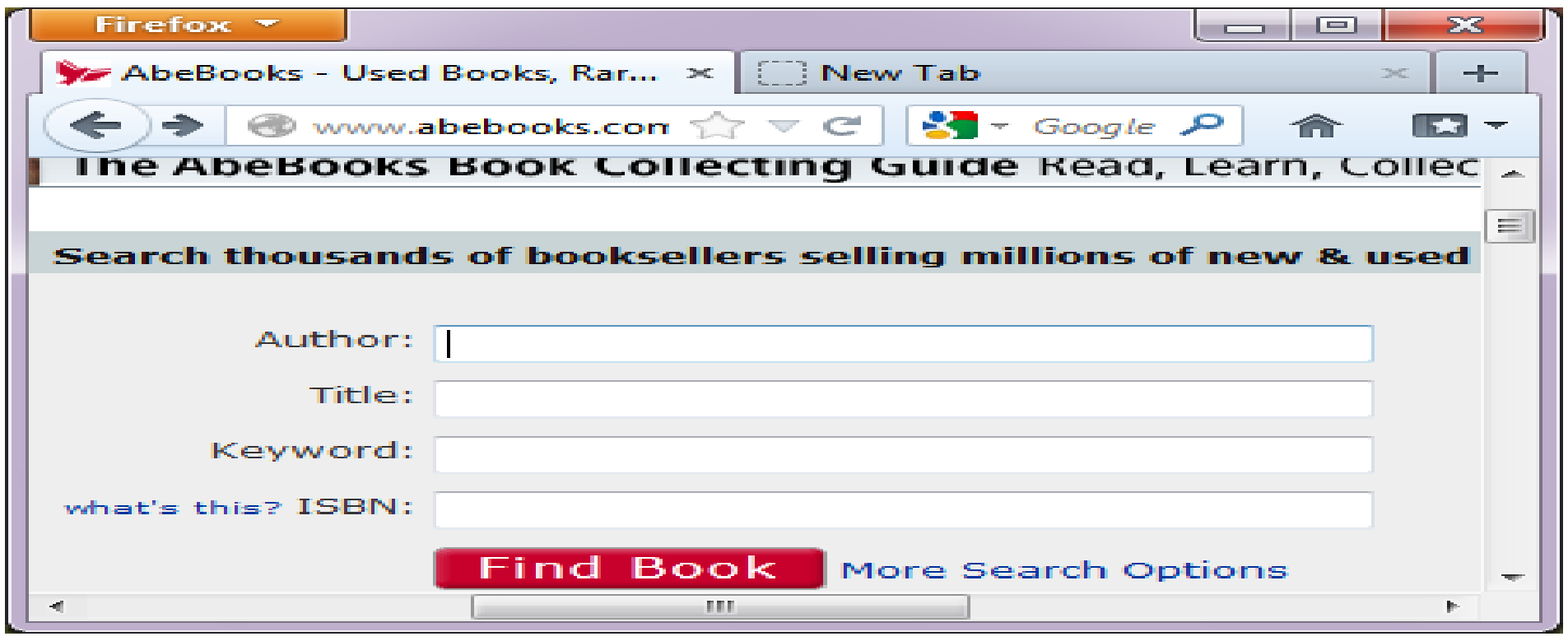

**Fig. 8** Multi-attribute Search form.

The proposed DBSHWC (domain based semantic hidden web crawler) automatically fill forms on various websites by filling form using task specific database. This crawler will be initialized with some seed URLs. Task Specific database is generated described above as shown in Fig. 9.

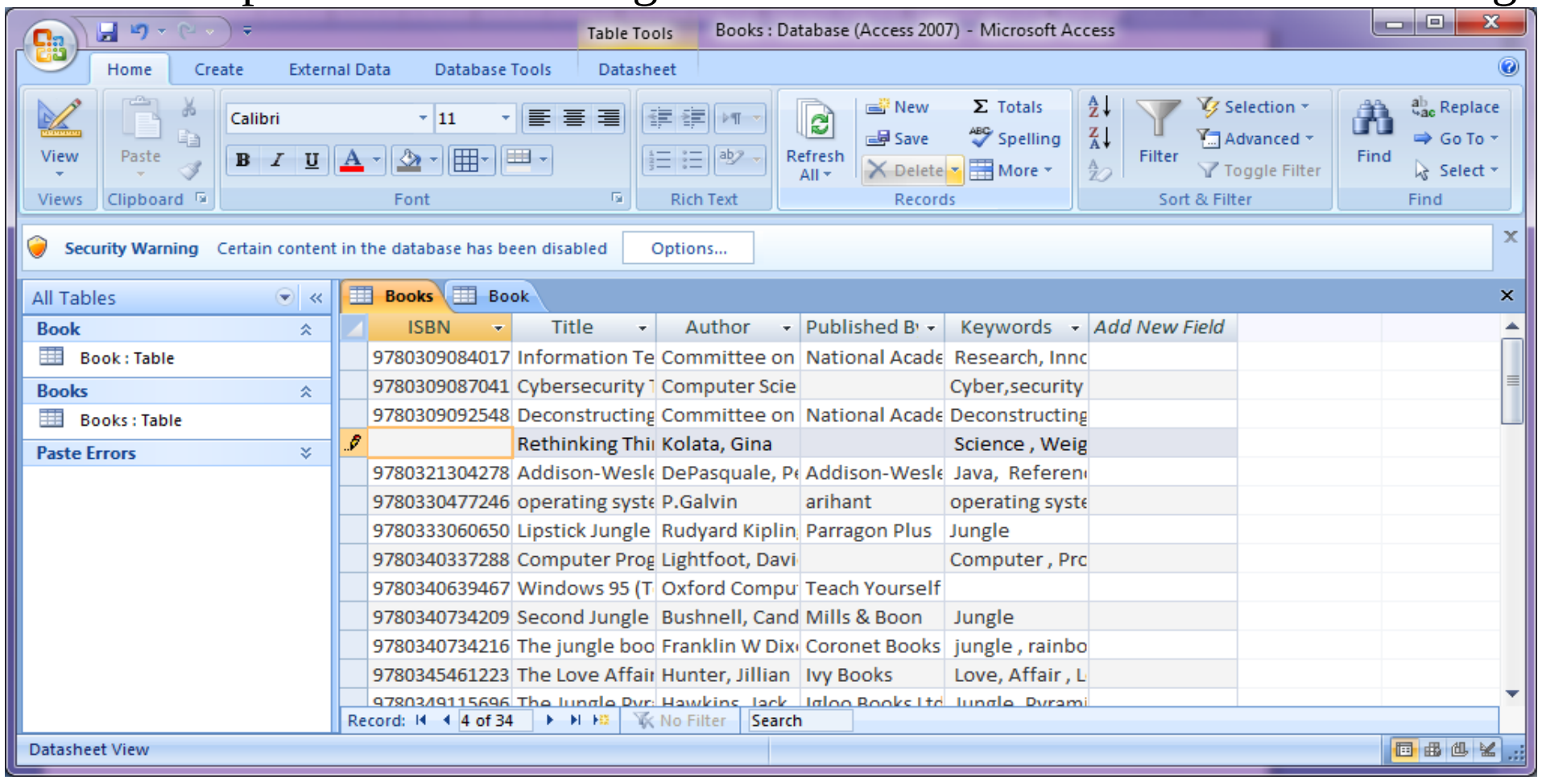

Fig. 9 Task Specific Database.

Fig. 10 shows automatic form filling process where the HWC take the values from the database by matching various labels of the fields present on the form page with Database Labels/Values.

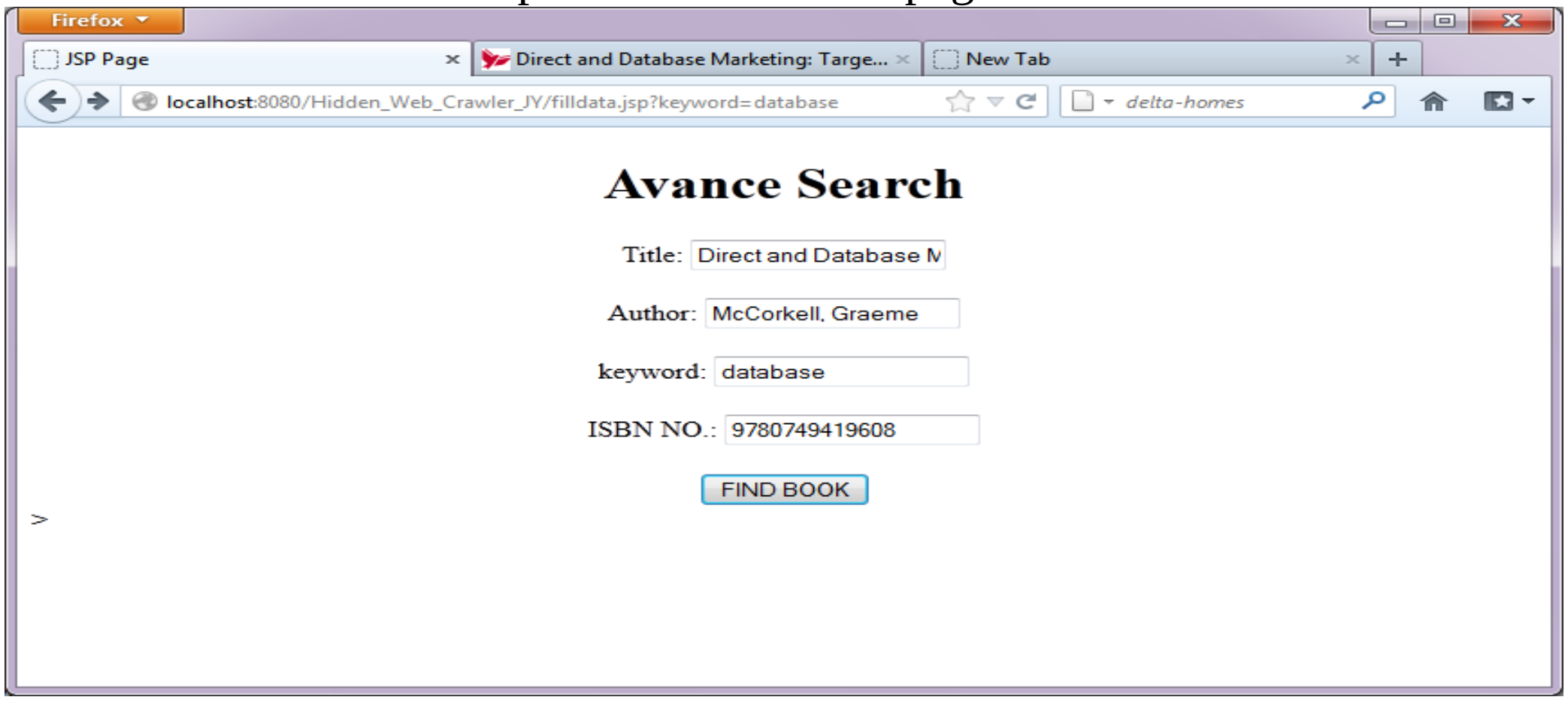

Fig. 10 Automatic Filled Form.

After filling forms and result processing, a Hidden Web Repository is developed which will have all hidden web pages of book domain. These hidden web pages are generated by filling search interface on book domain websites by this proposed crawler. Hidden Web data repository is shown in Fig. 11.

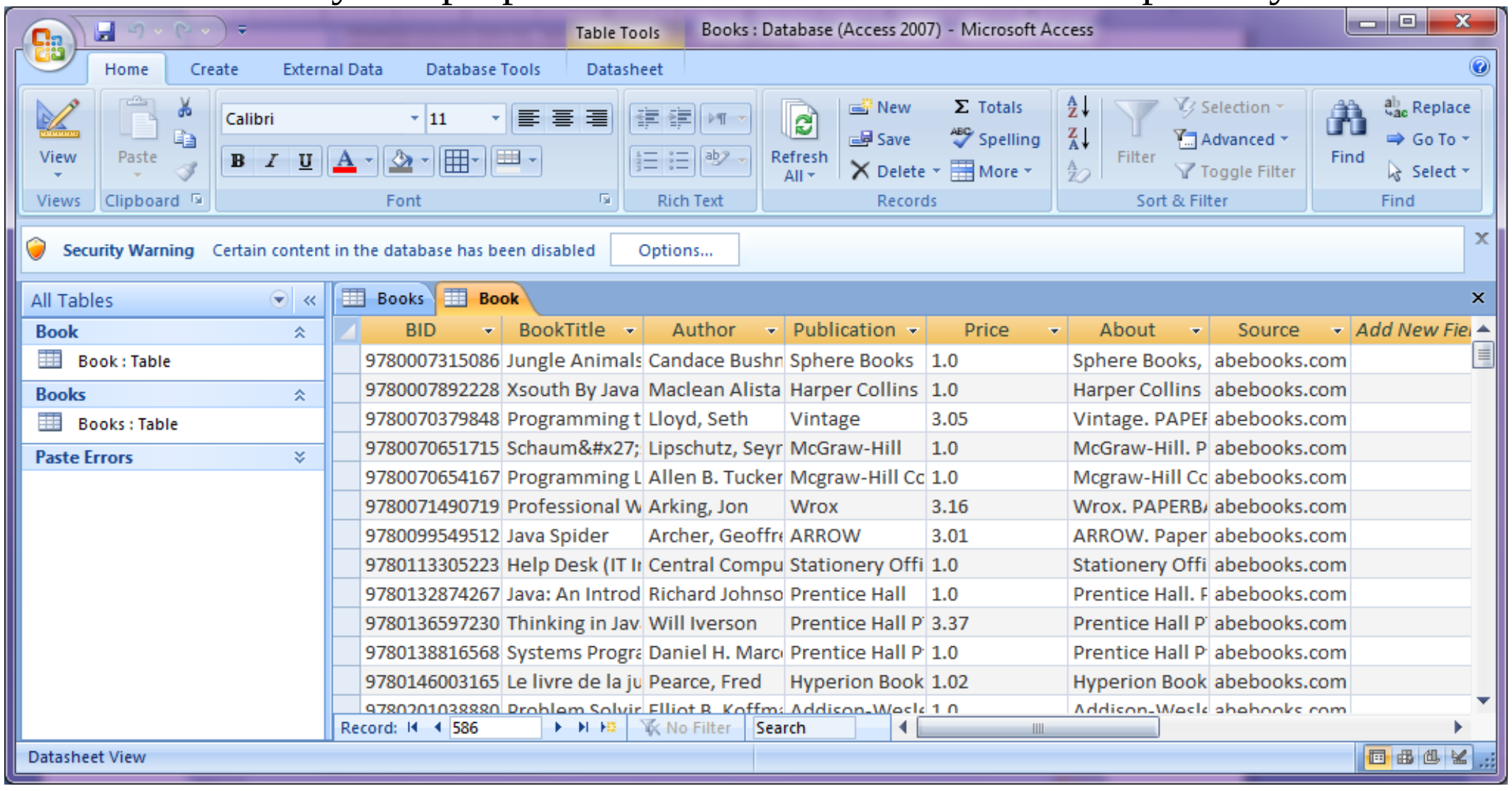

| BID | BookTitle | Author | Publication | Price | About | Source | Add New Fie |
|---|---|---|---|---|---|---|---|
| 9780007315086 | Jungle Animals | Candace Bushn | Sphere Books | 1.0 | Sphere Books, | abebooks.com | |
| 9780007892228 | Xsouth By Java | Maclean Alista | Harper Collins | 1.0 | Harper Collins | abebooks.com | |
| 9780070379848 | Programming t | Lloyd, Seth | Vintage | 3.05 | Vintage. PAPEF | abebooks.com | |
| 9780070651715 | Schaum' | Lipschutz, Seyr | McGraw-Hill | 1.0 | McGraw-Hill. P | abebooks.com | |
| 9780070654167 | Programming L | Allen B. Tucker | Mcgraw-Hill Cc | 1.0 | Mcgraw-Hill Cc | abebooks.com | |
| 9780071490719 | Professional W | Arking, Jon | Wrox | 3.16 | Wrox. PAPERB/ | abebooks.com | |
| 9780099549512 | Java Spider | Archer, Geoffr | ARROW | 3.01 | ARROW. Paper | abebooks.com | |
| 9780113305223 | Help Desk (IT I | Central Compu | Stationery Offi | 1.0 | Stationery Offi | abebooks.com | |
| 9780132874267 | Java: An Introd | Richard Johnso | Prentice Hall | 1.0 | Prentice Hall. F | abebooks.com | |
| 9780136597230 | Thinking in Jav | Will Iverson | Prentice Hall P | 3.37 | Prentice Hall P | abebooks.com | |
| 9780138816568 | Systems Progr | Daniel H. Marc | Prentice Hall P | 1.0 | Prentice Hall P | abebooks.com | |
| 9780146003165 | Le livre de la ju | Pearce, Fred | Hyperion Book | 1.02 | Hyperion Book | abebooks.com | |
| 9780201038880 | Problem Solvir | Elliot B. Koffm | Addison-Wesl | 1.0 | Addison-Wesl | abebooks.com | |

Fig. 11. Hidden Web Data Repository.

A global interface for user is developed which will take input from user and will give corresponding results from Hidden Web data Repository.

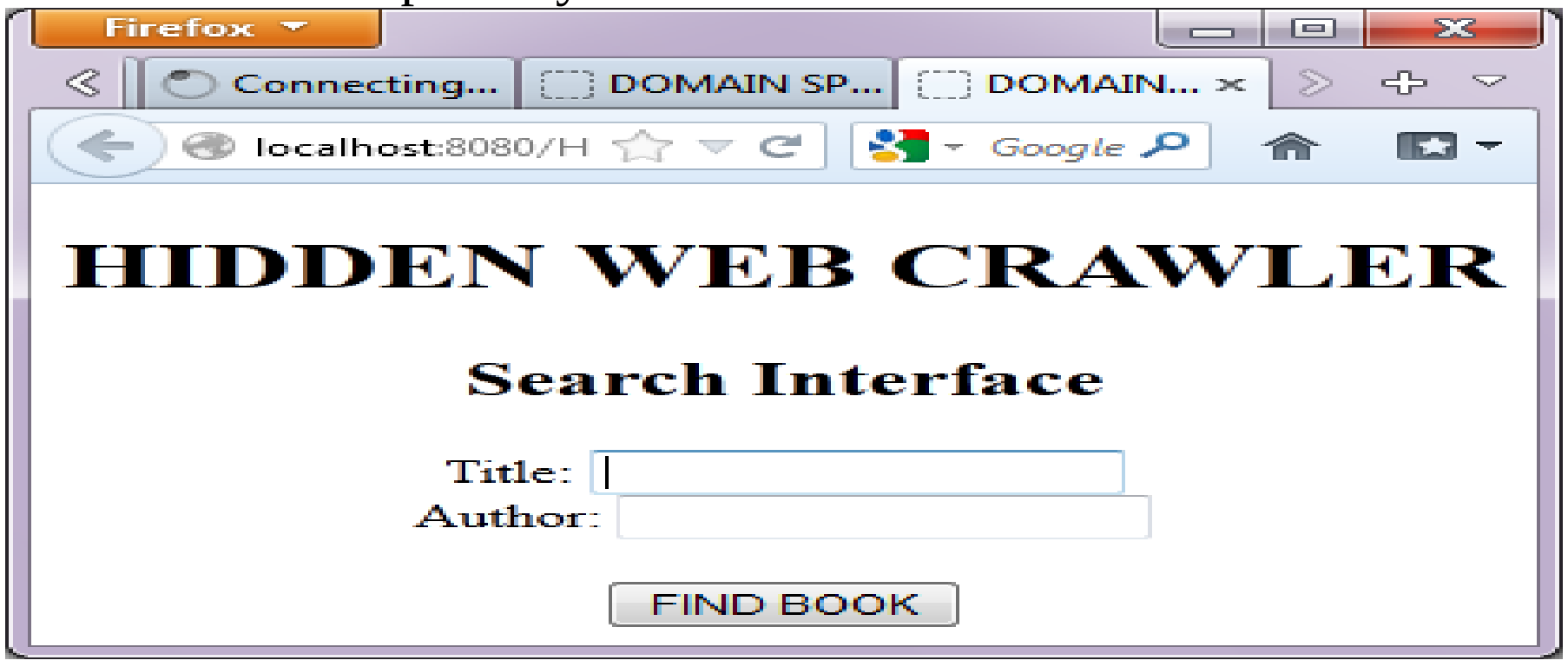

Fig. 12. New Designed hidden Web User Interface.

User query is converted into sql query to retrieve corresponding results . For e.g. If user entered title and author, then Sql query will be formulated as-

Select * from TableName WHERE BookTitle LIKE'%"+title+"%' AND Author LIKE '%"+auth+"%'

Results are shown to users as shown in fig. 13 .

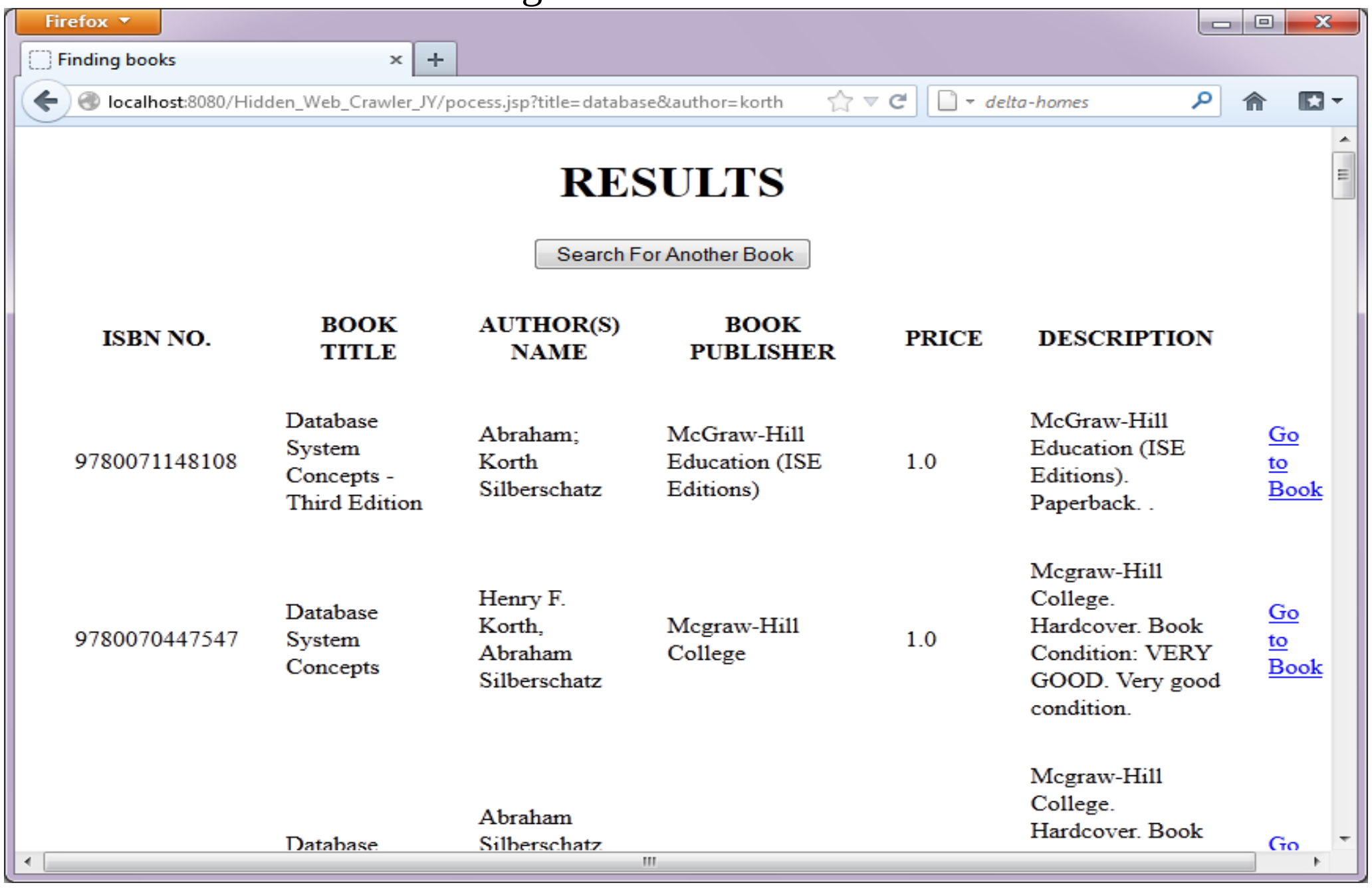

Fig. 13 Results To user from Hidden Data Repository

**Result evaluation**

The following table shows the results of proposed architecture. Where the Valid (correct) pages are computed by the matching them with user queries and satisfaction from the result retrieved. The percentage of total no. of valid pages can be calculated as:

$$\%\ \text{of Valid Pages} = \frac{\text{Number of Valid Pages}}{\text{Total number of pages Retrieved.}}$$

| | Traditional Hidden Web Crawler | Domain Based Semantic Hidden Web Crawler |
|---|---|---|
| No. Of Website Visited | 20 | 20 |
| No. Of Forms Filled | 36 | 48 |
| Total No. Of Pages | 428 | 528 |
| Total No. Of Valid Pages | 291 | 486 |

Table 2 Comparison of results.

Graph 1 shows the graphical representation of comparison of total number of retrieved web pages and number of valid result web pages by traditional hidden web crawler and proposed domain based semantic hidden web crawler.

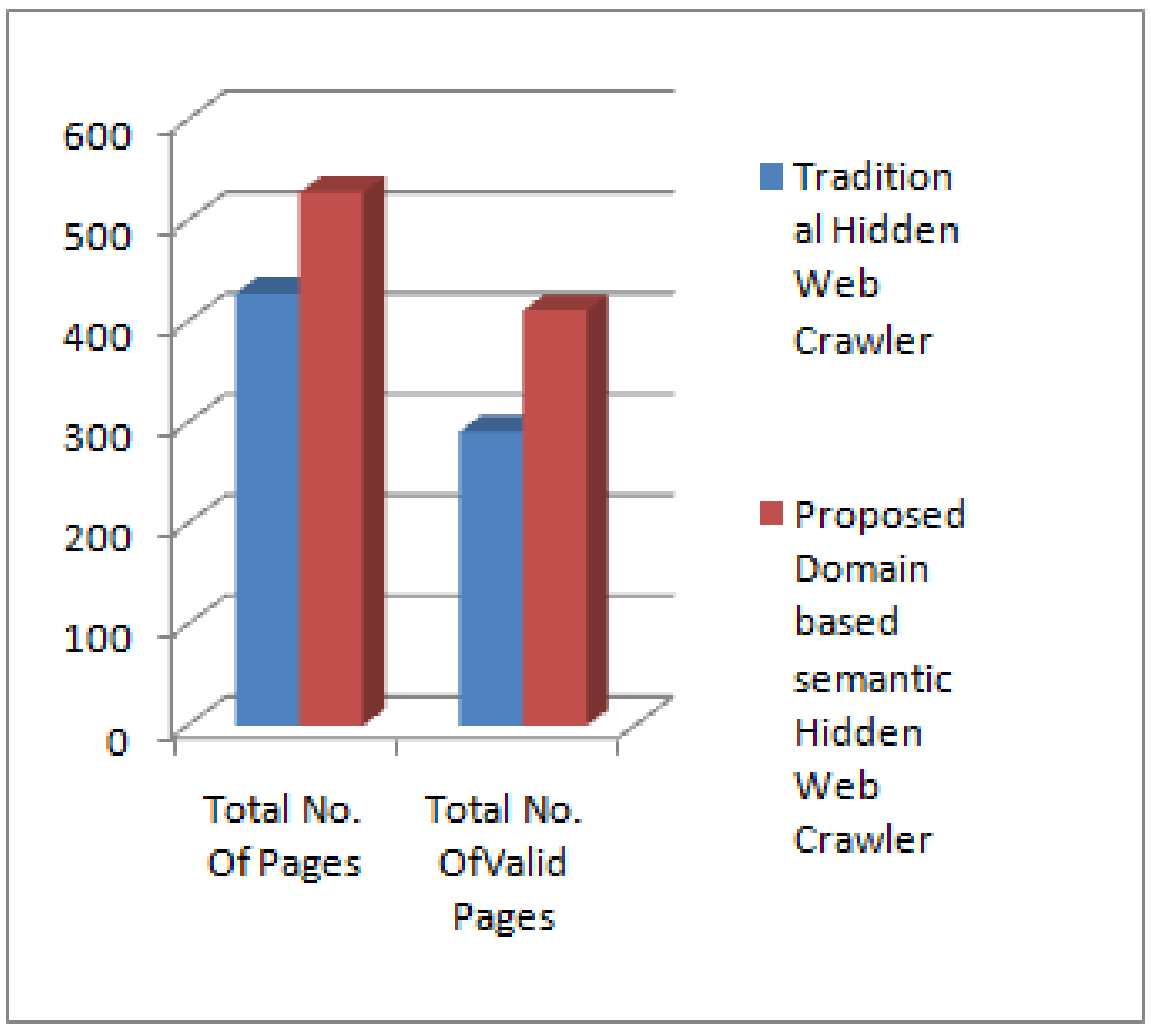

Fig. 14 Graph Showing Total pages and Valid pages.

Graph 2 shows the graphical representation of comparison of number of redundancy and duplicity in retrieved result web pages by traditional hidden web crawler and proposed domain based semantic hidden web crawler.

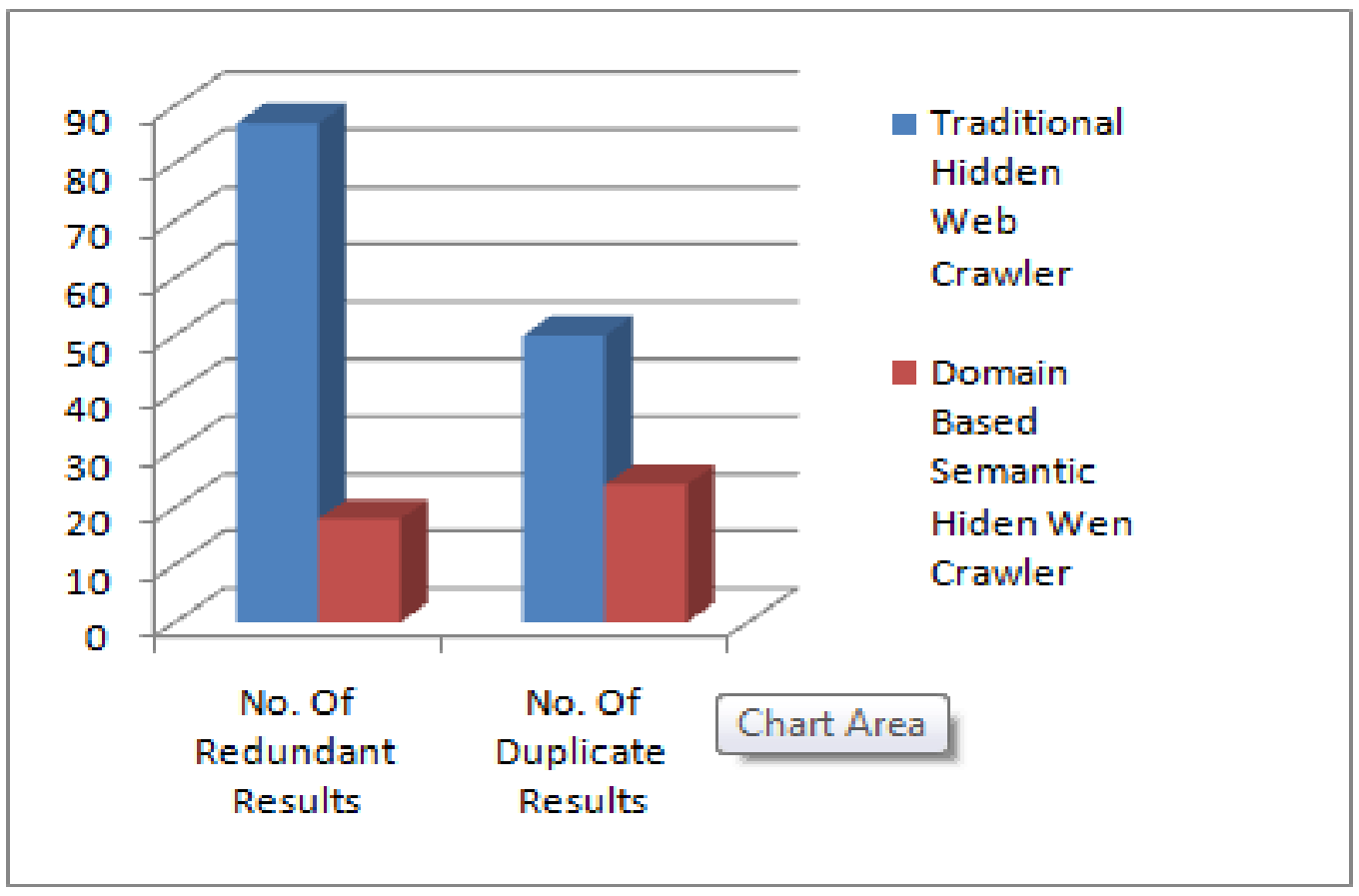

Fig. 10. Comparison of Redundant and duplicate results

From the results it can be observed that there is improvement in the percentage of forms filled and no. of valid pages retrieved in comparison with traditional HW crawlers.

## 6. CONCLUSION AND FUTURE SCOPE

Hidden Web data is now becoming highly important so extraction of hidden data is quite relevant. Hidden data cannot be directly extracted so we need special technique which can extract hidden data by filling search forms and generating relevant queries for the hidden web database. This paper entitled 'domain based semantic hidden web crawler' provides a simple and efficient technique to extract data from hidden web database of interested domains. It will differentiate databases on basis of domains and thus provides better relevant results for the user queries.

This paper provide for extraction of hidden web data, further for more better results we also need indexing and ranking of hidden web data.